\documentclass[11pt,twoside]{article}
\usepackage{asp2010}

\resetcounters

\begin{document}

\title{Inner polar gaseous disks: incidence, ages, possible origin}
\author{O.~K.~Sil'chenko$^1$ and A.~V.~Moiseev$^2$
\affil{$^1$Sternberg Astronomical Institute of the Lomonosov Moscow State University,
University av. 13, 119991 Moscow, Russia}
\affil{$^2$Special Astrophysical Observatory, Russian Academy of Sciences, 357147 Nizhnii Arkhyz, Russia}}

\begin{abstract}
We review our current knowledge about a particular case of decoupled gas kinematics --
inner ionized-gas polar disks. Though more difficult to be noticed, they seem to be more
numerous than their large-scale counterparts; our recent estimates imply about 10\%\ of
early-type disk galaxies to be hosts of inner polar disks. Since in the most cases the
kinematics of the inner polar gaseous disks is decoupled from the kinematics of the outer
large-scale gaseous disks and since they nested around very old stellar nuclei, we speculate
that the inner polar disks may be  relic of very early events of external gas accretion several 
Gyr ago. Such view is in agreement with our new paradigm of disk galaxies evolution. 
\end{abstract}

\section{Introduction}

Among gas subsystems with decoupled kinematics, a particular interest is inspired by
polar rings/disks. Firstly, they are beautiful, secondly, they seem to be stable
over many dynamic times, and thirdly, they imply certainly accretion of external gas from highly
inclined orbits. \textbf{Inner} polar gaseous disks are less spectacular than large-scale 
polar rings; however
they may be even more numerous though difficult to be detected against the bright bulge background
in early-type disk galaxies.

We note that the first evidence of existence of circumnuclear gas on polar orbits in the 
literature was presented by \citet*{Rubin1977} in their interpretation of the large line-of-sight 
velocity gradient along minor axis in NGC~3672. Further \citet*{bettoni90} have claimed inner polar 
gaseous disk in  the  southern ringed lenticular galaxy NGC~2217. By studying it through long-slit 
spectroscopy, \citet{bettoni90} found visible gas counter-rotation in some slit orientations (not all). 
Their geometrical scheme for the center of NGC~2217 demonstrated clearly that the ionized-gas disk had 
to be warped in such a way that in the very center it occupied the polar plane orthogonal
to the bar major axis. Later we found inner polar disks in unbarred early-type spiral
galaxies NGC~2841 \citep*{n2841pr} and NGC~7217 \citep{n7217pr} by obtaining two-dimensional
velocity fields for the ionized gas and for the stellar component with the integral-field
unit of the 6-m telescope, Multi-Pupil Fiber Spectrograph (MPFS).
The outer neutral hydrogen in both spirals is confined
to their main symmetry planes and rotates normally. It was a puzzle how a small amount of
polar-orbiting gas could reach the circumnuclear regions without colliding with the main
gaseous disks.

Now a few dozens of inner polar gaseous disks/rings are known. Their samples were presented
earlier by \citet{corsini03} and \citet*{moiseev_malta}; the latest statistics based on the data
for 47 inner polar disks collected over literature is published by \citet{mois_rev},
and here we review briefly some incidence properties.

\section{Incidence}

When we analyze all the cases with the inner gaseous disks inclined to the galactic symmetry
planes by more than 45\deg, we find that the inclinations of such disks tend strongly to the
strictly polar orientation: about two thirds of all such disks are inclined by $\ga 80$\deg.
This is consistent with theoretical claims about stability of the strictly polar orientations
and instability of the disks inclined by intermediate angles; the latters would precess until
they occupy the polar or co-planar orientation.

The inner polar disks -- as well as the large-scale ones -- prefer to inhabit early-type
galaxies. However while large-scale polar rings are seen mostly around gas-poor E/S0 galaxies --
about a factor of 3 more often than around spirals, -- and it can be explained by them
devoiding the hosts with large-scale coplanar gaseous disks \citep{resh11}, the inner polar disks
are found in Sa--Sc spiral galaxies in one third of all cases, and large-scale coplanar
gaseous disks do not prevent their appearance (see the above mentioned examples of NGC~2841 and
NGC~7217); even a few cases are known to be found in very late-type dwarfs. The typical size
(radius) of an inner polar disk is 0.2--2 kpc; the lower limit is perhaps defined by our
restricted spatial resolution. If to consider inner polar disks together with the
large-scale relatives, a continuous sequence in their sizes normalized by a galaxy diameter is 
observed with a gap at the  size $\sim0.5\,D_{25}$. This bimodal distribution can be explained by
different agents of stability for polar structures: while the external structures are stabilized
by the spheroidal (or even triaxial) potential of halo, the inner disks are usually settled well 
within the bulge-dominated area \citep{SmirnovaMoiseev2013}.  In any case, the presence of 
embedding stabilizing potential
is important. Is it crucial that this three-dimensional potential has to be also triaxial 
as in NGC~2217 \citep{bettoni90}? \citet{mois_rev} presents the following statistics:
among 40 galaxies with the inner polar disks which have the morphological type S0 and later
there are 17 galaxies with bars or triaxial bulges. This gives us the fraction of barred
galaxies among galaxies with the inner polar disks, only 43\%$\pm$8\%, completely consistent
with the fraction of barred and/or triaxial-bulge galaxies among all disk galaxies, 45\%
\citep*{aguerri09}.

The list of all known till 2012 inner polar disks by \citet{mois_rev} cannot be used
to estimate how often the phenomenon can be met: the sample of the hosts of the inner
polar disks listed there is quite inhomogeneous. To estimate the inner polar disk
incidence, we have used the data of the recent integral-field spectroscopic survey
ATLAS-3D \citep{atlas3d_1}. The ATLAS-3D sample is volume-limited one and includes
60 elliptical galaxies and 200 lenticular galaxies (if we classify NGC~2768 as S0).
We have taken the raw science and calibration frames from the open Isaac Newton Group
Archive of the Cambridge Astronomical Data Center and have calculated the stellar
and ionized-gas line-of-sight velocity fields. Then the orientation of the rotation
planes for both components in every galaxy was determined by fitting a circular-rotation
model, and the angles between the rotation planes of the stellar and gaseous components
were calculated  by using the formula (1) from \citet{mois_rev}. Among 200 S0 galaxies
of the ATLAS-3D volume-limited sample, we have found 8 new inner polar gaseous disks
with the inclination to the stellar rotation planes by more than 50\deg\
(taken into account both solutions of the equation (1) of \citet{mois_rev}, 
because we don't know  which side of the  is nearest to the observer);
12 inner polar disks in the S0 galaxies of the ATLAS-3D sample have been already listed
in \citet{mois_rev}. Having in total 20 inner polar disks in S0 galaxies of the
ATLAS-3D volume-limited sample, we conclude that nearby lenticular galaxies have
inner polar disks in 10\%\ of all cases. Our estimate refers to the totality of S0 galaxies
over {\it all} types of environments. This incidence of the inner polar disks in
the early-type disk galaxies, 10\%, exceeds greatly the frequence of the large-scale
polar rings, 0.1--0.4\% \citep{resh11}.

\begin{figure*}[!ht]
\centerline{
\resizebox{0.8\hsize}{!}{\includegraphics{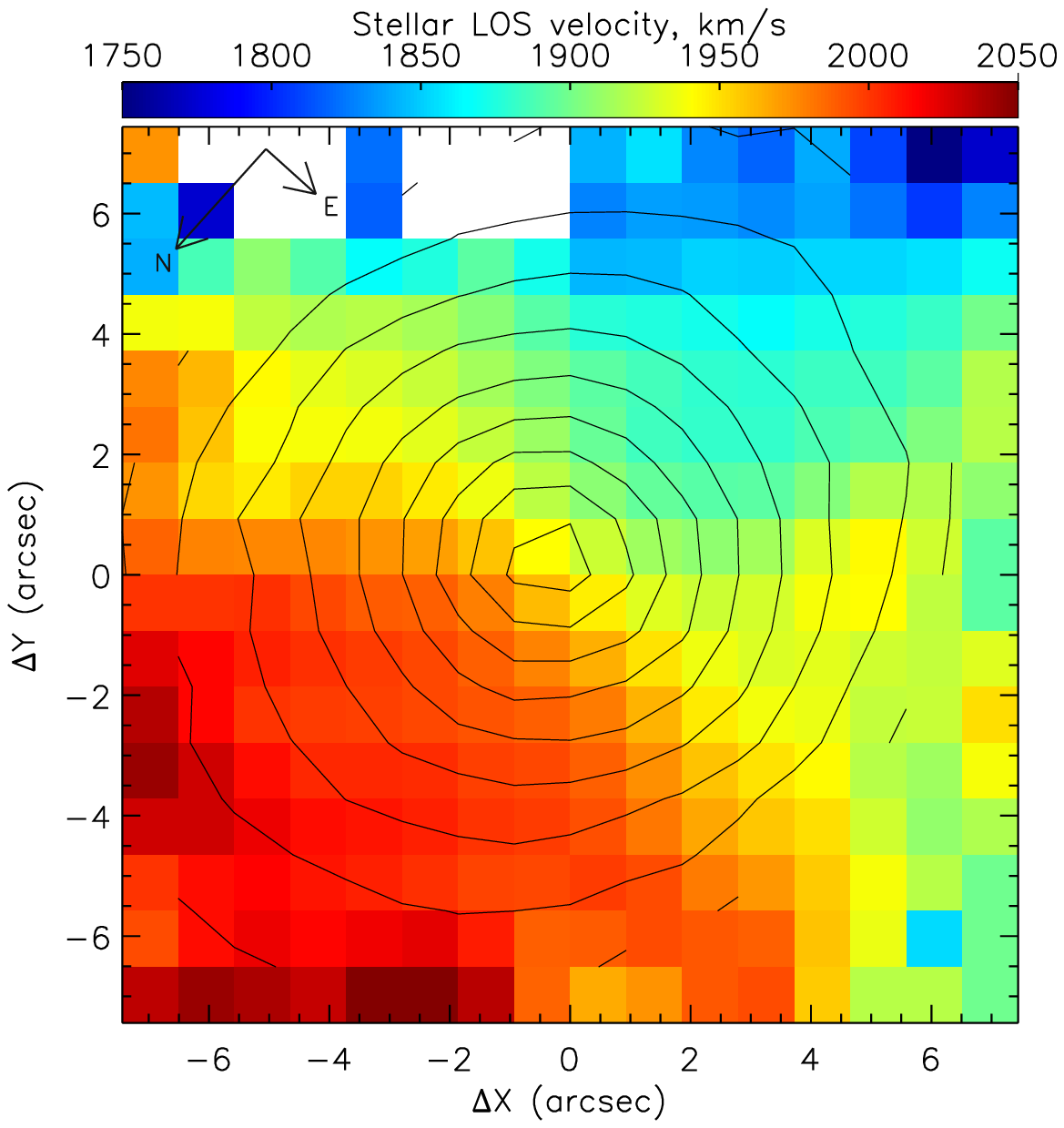}
\includegraphics{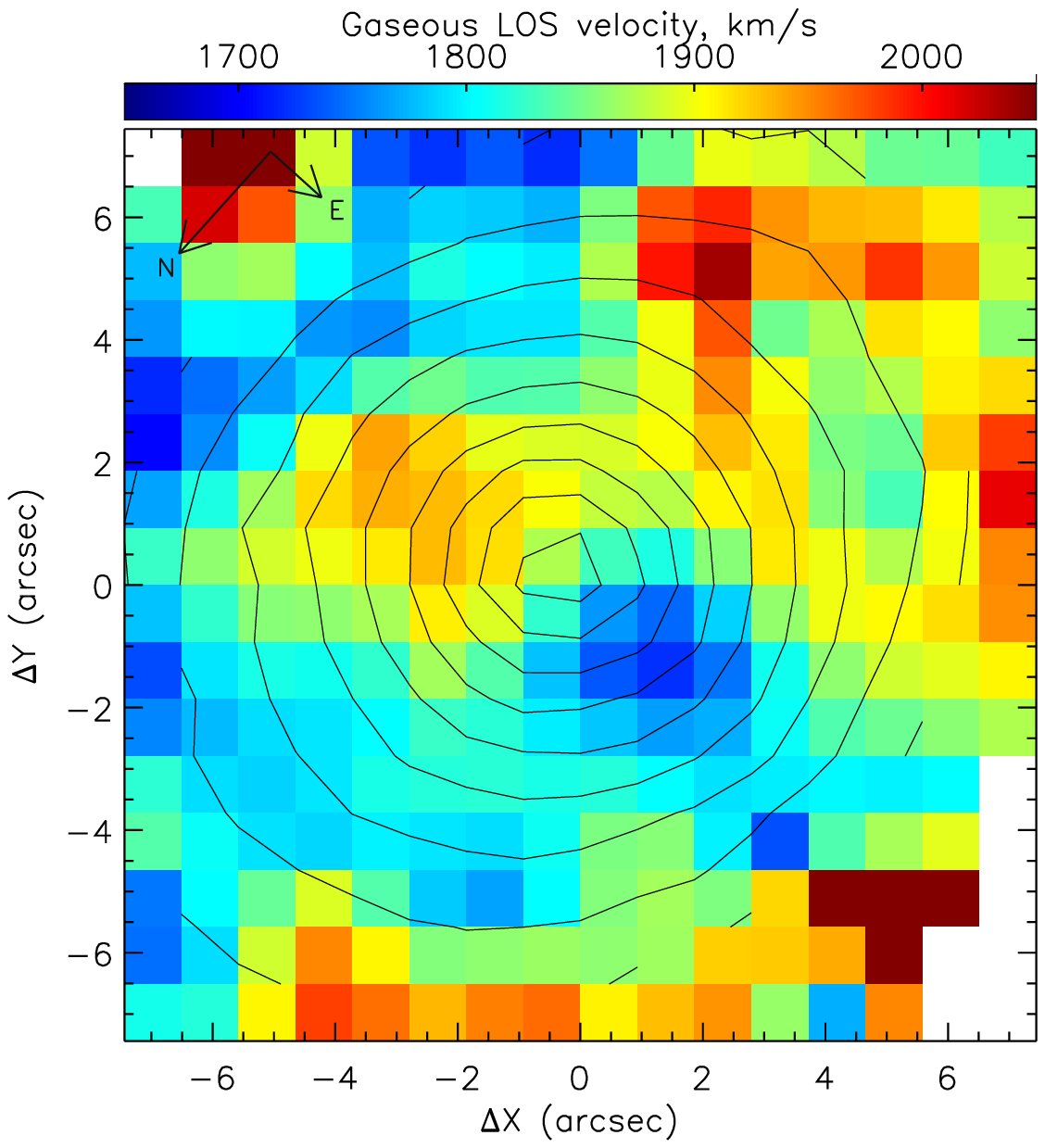}}
}
\centerline{
\includegraphics[width=0.8\textwidth]{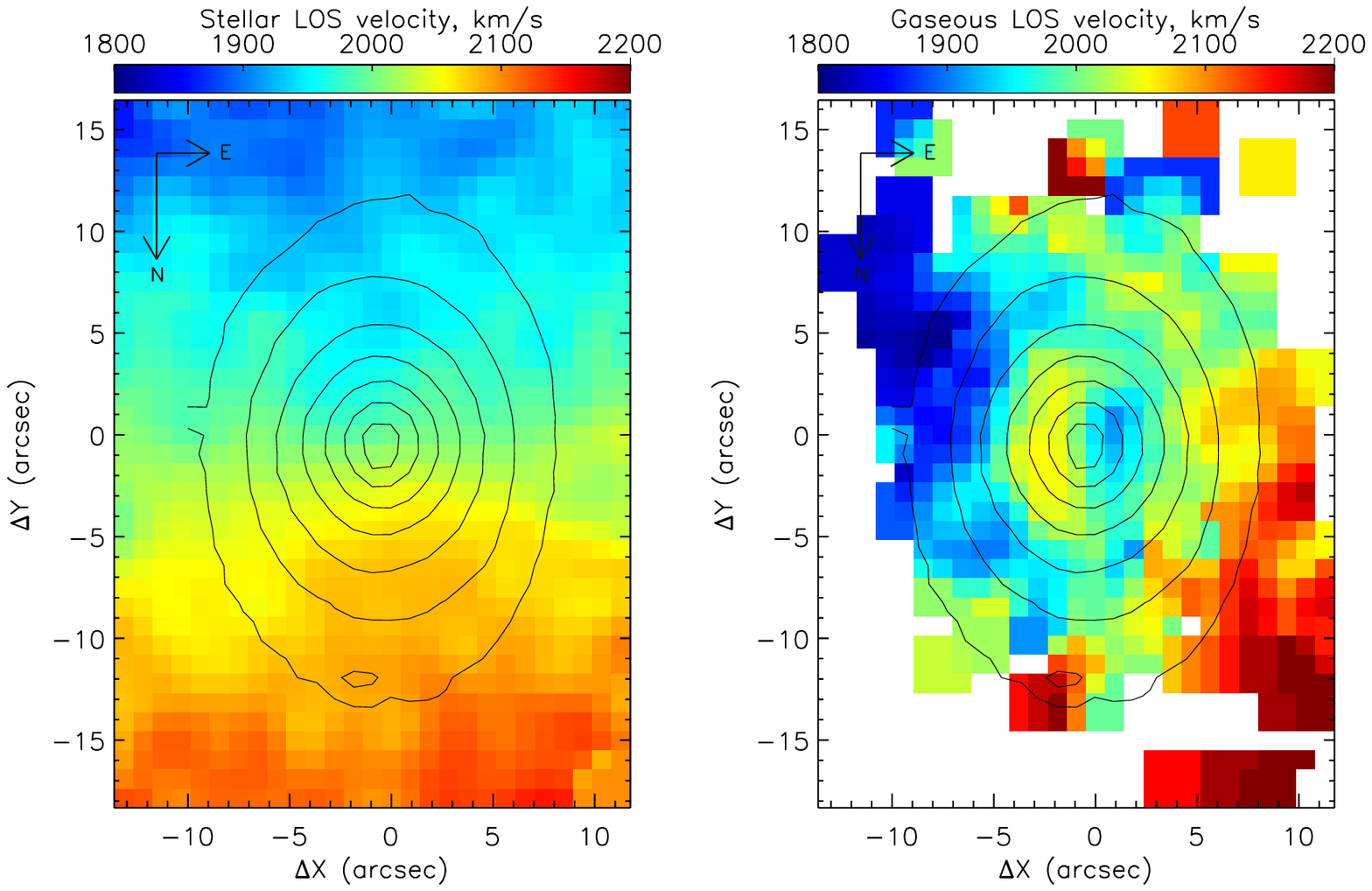}
}
\caption{The line-of-sight velocity fields for the stellar and ionized-gas components
in the lenticular galaxy NGC~2962: the upper row presents the data from the  MPFS
of the Russian 6-m telescope, the bottom raw -- our reduction of the SAURON data.}
\label{n2962}
\end{figure*}

Figure~\ref{n2962} shows a nice example of the newly discovered inner polar disk in
the lenticular galaxy NGC~2962 -- a member of the ATLAS-3D volume-limited sample. We have
observed this galaxy earlier  at the  Russian 6-m telescope with the integral-field spectrograph  MPFS which field of view was $16\arcsec \times 16\arcsec$, and in the very center, inside $R=5\arcsec$, we saw a compact, fastly rotating, nearly edge-on polar gaseous disk. But with the larger field
of view of the SAURON, $41\arcsec \times 33\arcsec$, we are now seeing a switch of the gas
rotation sense at $R\approx 7\arcsec -10\arcsec$: the galaxy possesses two nested
polar gaseous disks counterrotating each other (Fig.~\ref{n2962}).

\section{Origin}

\subsection{Is the polar momentum inner or external?}

This question may seem to sound strange: if a main baryonic component, stars which are
formed from the own gas of the galaxy, rotates in the galactic disk symmetry plane,
how may the polar gas be of local origin? Meanwhile there are intrinsic secular evolution
mechanisms that produce strongly inclined gaseous disks in the very center of a galaxy,
and one of them had been revealed by simulations of \citet{friedli_benz}. By tracing
dynamical evolution of {\it initially retrograde} gas in the disk of an isolated barred
galaxy, \citet{friedli_benz} have found that after about 2~Gyr of angular momentum exchange
with the stellar bar the gas inside a few hundred parsec comes to a strongly inclined plane 
due to vertical instabilities. Since retrograde motions of stars are always present in the barred
potential \citep{pfenniger}, and since stars drop gas during their evolution, in principle
the inner polar gaseous disks may form in barred galaxies without outer donor contribution.
Indeed, we have found several cases when the presence of the inner polar disk in the
very center is accompanied by the presence of counterrotating gas in the more outer
disk -- e.g. in NGC~7280 \citep{n7280,sil05}.
But the presence of a bar is necessary. However, the statistics in the previous Section
does not show prevalence of barred galaxies among the hosts of inner polar disks: less
than a half of the hosts of inner polar disks reveal triaxiality of their inner stellar
structures. So we are now inclined to the hypothesis of the external gas accretion as
the dominant mechanism of inner polar disk formation.

\begin{figure*}[!ht]
\centerline{
\includegraphics[width=0.8\textwidth]{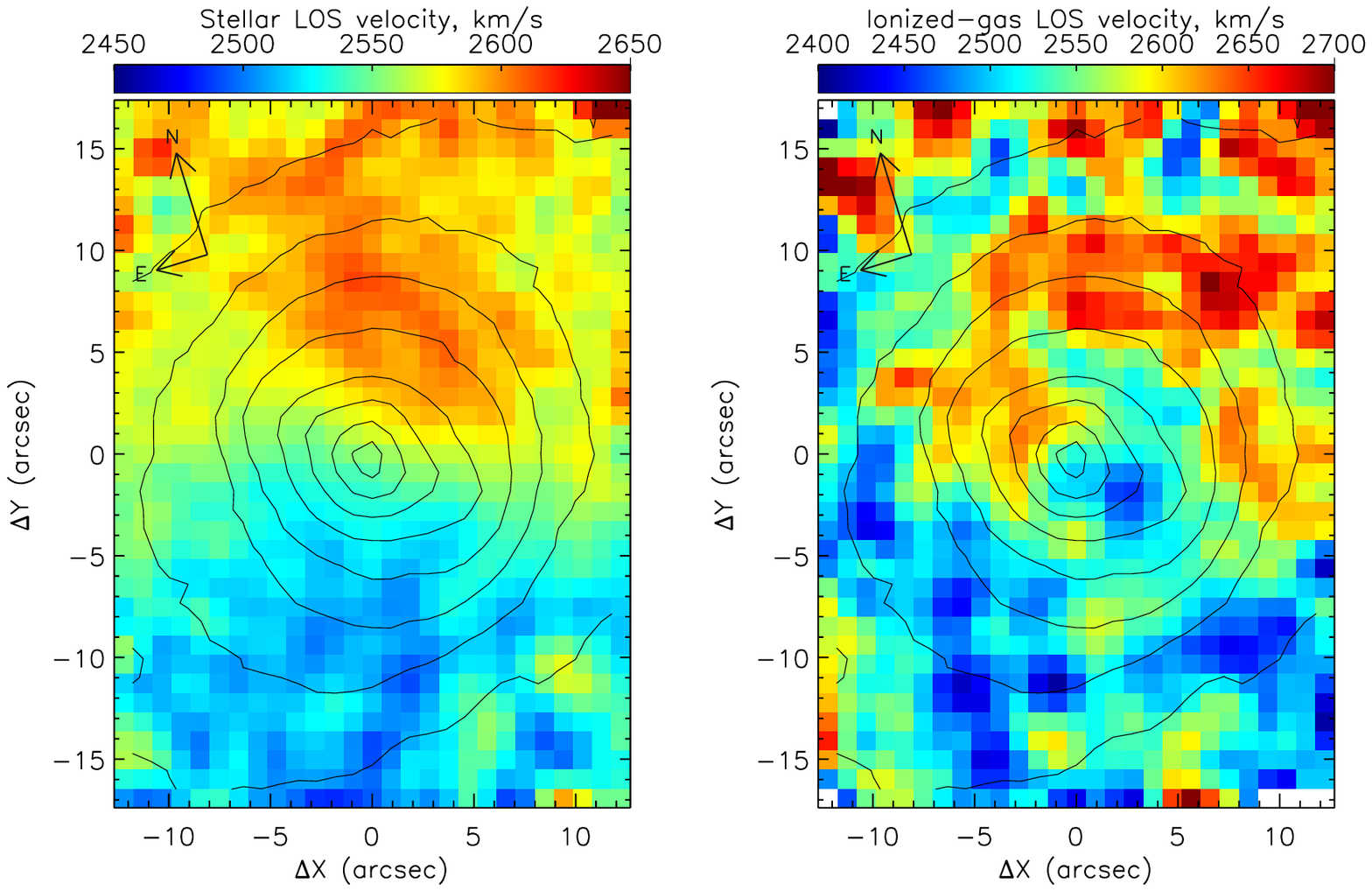}
}
\caption{The line-of-sight velocity fields for the stellar (left) and ionized-gas components
(right) for the spiral galaxy NGC~5850 from our reduction of the SAURON data.}
\label{n5850}
\end{figure*}

\subsection{How much gas can be in a polar orbit?}

To identify a source of gas accretion, we must estimate first of all typical amounts of
gas populating polar orbits. Here a lot of diversity is observed. In some cases the inner
polar ionized-gas disks have their extension into the very outer parts of galaxies when
they are observed at the 21cm line of the neutral hydrogen -- these are the cases, e.g., of
NGC~3414 (with the inner polar disk found by \citet{polardust}) or of NGC~7280 or of UGC~9519
mapped in the neutral hydrogen line by \citet{serra_h1}.
In the prototype of large-scale polar ring galaxies, NGC~2685, the inner ionized gas is
also polar \citep{sil_polar98}. In these cases the total mass of the polar gas can be as
large as $10^8 -10^9$ solar masses, and the $M(\mbox{HI})/L_K$ ratios resemble those
of spiral galaxies \citep{serra_h1}. In the volume-limited S0-galaxy sample from ATLAS-3D
\citep{atlas3d_1} about one third of all galaxies with the inner polar ionized-gas disks
have polar neutral-hydrogen outer extension. However many galaxies have inner {\it polar}
ionized-gas component and outer {\it coplanar} neutral-hydrogen disk; and they are sometimes
also rather gas-rich but their main gaseous components are confined to the galaxy symmetry planes.
Among lenticular galaxies, we can mention NGC~2962 where \citet{grossi09}
have found $1.1\times 10^9$~M\sun\ of neutral hydrogen in a disk coplanar to the
stellar one but extending much farther from the center. And certainly even more such
cases can be found among spiral galaxies with the inner ionized-gas polar disks.
An inner ionized-gas polar disk was found in a barred spiral, SB(r)b, galaxy
NGC~5850 by \citet*{moiseev04}; the stellar and gaseous rotations were compared over
the $16\arcsec \times 16\arcsec$ field of view of the 6-m telescope IFU MPFS.
Now we have calculated larger stellar and gaseous velocity fields by using the archival 
SAURON data (Fig.~\ref{n5850}). One can immediately see from Fig.~\ref{n5850} that
the sense of the gas rotation changes at the radius of 7\arcsec--10\arcsec\ (1.3--1.8 kpc);
the more outer ionized gas rotates together with the stars. And the same orientation
of the rotation plane is demonstrated by all the $2\times 10^9$ solar masses of neutral
hydrogen measured in NGC~5850 by \citet*{higdon}. The same  patterns of stellar and ionized gas  
circumnuclear kinematics were also presented recently in the paper by \citet{Bremer2013}, which 
is based on VLT observations with the VIMOS IFU. The better spatial resolution 
\citep[comparing with the early observations by][]{moiseev04} has allowed to calculate precisely 
the kinematic orientation parameters in the inner disk velocity field. \citet{Bremer2013} claimed 
that the angle between the inner and outer disks planes is only $24\deg$, however 
the equation (1) from \citet{mois_rev} gives also the second solution -- $54\deg$, 
that corresponds to the case of strongly inclined inner gaseous disk.

\begin{figure*}
\centerline{
\includegraphics[width=0.5\textwidth]{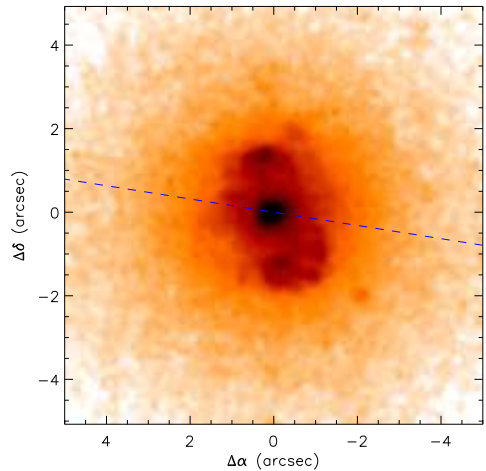}
}
\caption{The narrow-band emission-line (F658N {\it minus} F814W) image of the central part 
of NGC~7217 obtained with the camera ACS/HST. The dashed line shows the line of nodes
of the galactic stellar disk.}
\label{n7217}
\end{figure*}

\subsection{NGC~7217}

An interesting case of a spiral galaxy with
the inner polar ionized-gas disk having the radius of only 350 pc \citep{n7217pr} is
represented by an isolated Sab galaxy NGC~7217; here we show the recent HST image of
the central part of the galaxy (Fig.~\ref{n7217}) where the inner ionized-gas polar
disk can be seen `by eye' in the narrow photometric band centered onto the emission
lines H$\alpha +$[NII]. Its neutral hydrogen disk, $0.7\times 10^9$~M\sun, extending
to $R\approx 8$ kpc, is coplanar to the stellar disk and rotates just as the stars;
at the outer edge of this disk intense star formation in a ring is observed
though the visible gas density is below the gravitational stability threshold \citep{n7217_h1}. 
Recently we have studied the origin of the complex structure of NGC~7217 in detail
\citep{last7217}, and here we discuss this galaxy as a pure key point revealing possible
formation mechanisms of the inner polar disks.

Photometric structure of NGC~7217 can be described as three-tiered: we \citep{n7217pr}
have separated three exponential segments in its surface-brightness radial profile.
The innermost segment seen only at $R<10\arcsec$ (0.8~kpc) may be a pseudobulge; then other
two segments represent an antitruncated disk. Our deep long-slit spectroscopic observations
\citep{last7217} having allowed to measure stellar rotation and line-of-sight velocity
dispersion (close to a vertical velocity dispersion because the galaxy is seen almost
face-on) as well as the properties of the stellar populations, have revealed prominent
differences in all respects between two exponential parts of the stellar disk.
Firstly, the inner part of the disk is substantially  thinner than the outer part,
and secondly, the mean age of the stellar population in the inner disk is 5~Gyr while
the stellar ages in the outer disk, even beyond the starforming ring, is very young, less
than 2~Gyr. The galaxy being an early-type spiral without a bar, possesses meantime
three rings of current star formation \citep{verdes95}. Interestingly, the age of the
nuclear stellar population, inside the circumnuclear starforming ring, is {\it very} old --
older than 10~Gyr. Obviously, despite violent processes of gas radial re-distribution
and external gas accretion betrayed by the inner polar disk presence,
the gas has never reached the very center of NGC~7217 for the last 10~Gyr.

Having in hands the detailed structure of NGC~7217 and evolutionary sequence of building elements
of this structure, we have tried to fit observational properties of NGC~7217 with the models
provided by on-line service GalMer \citep{galmer}. We have found that only at least {\it two}
independent gas-rich minor-merger events can provide a full list of properties: the inner polar
disk is formed by an accretion of a gas-rich dwarf from an inclined retrograde orbit, and
the outer flaring ringed starforming disk is shaped by merging a prograde-orbiting
satellite. The necessity of two minor mergers is due to the fact that minor merging
from a retrograde orbit gives an inclined inner gaseous disk but does not thicken the
large-scale stellar disk. The latter feature requires minor merging from a prograde orbit.
Since the star formation burst in the outer disk of NGC~7217 is very young, we conclude
that the minor merging from a retrograde orbit was the first event, and minor merging from
a prograde orbit was the last, quite recent one.

\section{Ages}

The large-scale outer polar rings may be stable in the polar state over a few Gyrs  
according to theoretical estimations \citep[e.g.][]{steiman-cameron} as well as to numerical 
simulations \citep{Snaith2012}. Stability of their circumnuclear counterparts is still an open 
question. However some observational evidences in favour of their very long living times also exist: 
just among lenticular galaxies with the inner polar disks we found {\it very} old stellar nuclei, 
$T>10$~Gyr \citep{polardust}, while over the full sample of nearby lenticular galaxies 
the typical ages of the stellar nuclei are 2--5~Gyr \citep{sil06,sil08}. 

The whole evolution of disk galaxies is governed by the regime of external gas accretion.
Recently we \citep{sil12} have proposed a scenario according to which all disk galaxies were
formed around $z\approx 2$ as lenticular galaxies, and only much later, at $z<1$, most of
them started smooth gas accretion and, after having formed thin dynamically cold stellar disks,
transformed into spirals. In the frame of this scenario, a natural epoch of forming
inner polar gaseous disk is very early stages of the accretion era. If the first accretion
event was from a highly inclined orbit, an inner polar long-living gaseous disk would
form {\it before} the main gas accretion in the galactic symmetry plane proceeded.
It is the way to obtain a stable system with mutually orthogonal nested gaseous disks;
and then inner polar disks would be relics of very early events of external gas accretion.

\acknowledgements We thank Enrica Iodice and the organizers for the interesting and inspiring 
conference and for the invitation to present this review. A.M. is  grateful to the non-profit 
`Dynasty' Foundation and to the RFBR grant 13-02-00416. This contribution makes use of data obtained 
from the Isaac Newton Group Archive which is maintained as part of the CASU Astronomical
Data Centre at the Institute of Astronomy, Cambridge. The ACS images of NGC 7217  were obtained  
from the Hubble Legacy Archive, which is a collaboration between the Space Telescope Science Institute 
(STScI/NASA), the Space Telescope European Coordinating Facility (ST-ECF/ESA) and the Canadian Astronomy 
Data Centre (CADC/NRC/CSA).

\bibliography{001osilchenko}

\end{document}